\documentstyle[12pt]{article}
\newcommand{\be}{\begin{equation}}
\newcommand{\ee}{\end{equation}}
\newcommand{\ba}{\begin{array}{rcl}}
\newcommand{\ea}{\end{array}}
\begin{document}
\begin{center}
{\Large\bf Quantum chaos in small quantum networks}\\
\medskip
Ilki~Kim and G\"{u}nter~Mahler\\
Institut f\"ur Theoretische Physik~I, Universit\"{a}t Stuttgart\\
Pfaffenwaldring 57, 70550 Stuttgart, Germany\\
phone: ++49-(0)711 685-5100, FAX: ++49-(0)711 685-4909\\
email: ikim@theo.physik.uni-stuttgart.de\\
\end{center}

\begin{abstract}
 We study a 2-spin quantum Turing architecture, in which discrete local
 rotations $\{\alpha_m\}$ of the Turing head spin alternate with quantum
 controlled NOT-operations. We show that a single chaotic
 parameter input $\{\alpha_m\}$ leads to a chaotic dynamics in the entire
 Hilbert space. The instability of periodic orbits on the Turing head
 and `chaos swapping' onto the Turing tape are demonstrated explicitly
 as well as exponential parameter sensitivity of the Bures metric.
\end{abstract}

\section{Introduction}

In recent years problems of quantum computing~(QC) and information processing 
have received increasing attention. To solve certain classes of problems 
in a potentially very powerful way, one tries to utilize in~QC the 
quantum-mechanical superposition principle and the (non-classical) 
entanglement \cite{SCH35}, an undertaking which, at the same time, should 
contribute to our 
basic understanding of quantum mechanics itself (see e.g. \cite{DUE98}). 
However, building a large-scale quantum computer remains an extremely 
difficult task. The major 
obstacle is the coupling of the quantum computer to the environment, which 
tends to destroy quantum-mechanical superpositions very rapidly. 
This effect is usually referred to as decoherence. 
Present-day technology does not yet support the realization of 
a practical quantum computer. On the other hand, there might be 
interesting small-scale physics in a pure quantum regime based on a few 
pseudo-spins (qubits), which are realizable right now. 

Chaotic behaviour as an exponential sensitivity to initial conditions has 
been well established in classical non-linear systems. 
The deterministic chaos, which occurs in 
non-dissipative systems, can typically be found starting from 
regular states as a function of some external control parameter. 
However, there seems to be no direct analogue to 
chaos in the quantum world: If two quantum states are initially almost 
identical (that is, their\linebreak
scalar product is very close to $1$), 
they will remain so forever, since the Hamiltonian evolution 
is a unitary mapping which preserves scalar products. 
According to this negative result, the semiclassical 
`quantum chaology' \cite{BER85} has been constrained to studying 
some quantum-mechanical `fingerprints of chaos' (like spectral properties), 
and non-trivial transitions from the quantum - to the classical domain and 
vice versa, following Bohr's correspondence 
principle (see e.g. \cite{FOR91,HAA91}). In addition, 
experimental progress in mesoscopic physics, e.g. the transport of electrons 
through so-called `chaotic quantum dots' \cite{KOU97}, has attracted a great 
deal of interest, the results of which give numerical evidence for weak 
chaos (indicated by level repulsion) \cite{SHE94,WAI98}. 

While most models of QC have been concerned with networks of quantum 
gates, which are reminiscent of classical integrated circuits, 
models based on quantum Turing machines~(QTM) \cite{BEN82,DEU85} have been 
described along different lines but have not given rise to much potential for 
future applications up to now. In both cases the complexity of 
the computation is characterized by sequences of unitary transformations 
(or the corresponding Hamiltonians~$\hat{H}$ acting during finite time 
interval steps). 
The study of quantum chaos based on quantum gate networks 
has so far been proposed e.g. by implementing quantum baker's map 
on a 3-qubit NMR quantum computer \cite{SCH98}, by realizing a 
quantum-mechanical delta-kicked harmonic oscillator or a harmonically driven 
oscillator in an ion trap \cite{GAR97,BER99}, and by showing 
quantum-mechanical localization of an ion in a trap \cite{GHA97,RIE99}. 
In all these cases some sort of sensitivity has been 
predicted with respect to parameters specifying the dynamics 
(e.g. the respective Hamiltonian). 

Here we explicitly describe an iterative map with a few qubits 
which, though based on standard 
gates, can be thought to be realized as a QTM architecture: 
Local transformations of the Turing head controlled by a Fibonacci-like 
sequence of rotation angles alternate with a quantum-controlled NOT-operation 
with a second spin on the Turing tape. 
This chaotic control can generate a chaotic quantum propagation in the 
`classical' regime \cite{BLU94}, defined here as one with the 
Turing head being restricted to an entanglement-free state sequence 
(`primitive')~\cite{KIM99}. It will then be shown that chaos 
on the Turing head can be found also for the 
quantum-mechanical superposition of those primitives, 
implying entanglement between head and tape as a genuine quantum feature 
(`non-classical regime'). Finally, 
due to this quantum correlation, we find a chaotic propagation even 
in the reduced subspace of the Turing tape (`chaos swapping'), 
and as a result also in the total network state $|\psi_n\rangle$. 
This behaviour should be contrasted with that of a regular QTM with its 
long-time revivals, which are absent in the `chaotic' QTM.

\section{Description of chaotically driven quantum Turing machine}

The quantum network \cite{MAH98} to be considered is composed 
of $N$ ($=M+1$) pseudo-spins 
$|p\rangle\!^{(\mu)};\,p=0,1;\,\mu=S,1,2,\cdots, M$ 
(Turing-head $S$, Turing-tape spins $1,2, \cdots, M$) so that its 
network state $|\psi\rangle$ lives in the {$2^{M+1}$-dimensional} 
Hilbert space spanned by the product 
wave-functions $|j^{(S)} k^{(1)}\cdots l^{(M)}\rangle = |jk \cdots l\rangle$. 
Correspondingly, 
any (unitary) network operator can be expanded as a sum of product operators. 
The latter may be based on the following $SU(2)$-generators
\begin{eqnarray}
\label{lambda_s}
\hat{\lambda}_{1}^{(\mu)} &=& \hat{P}_{01}^{(\mu)} + \hat{P}_{10}^{(\mu)}
\,,\;\;\;\;
\hat{\lambda}_{2}^{(\mu)} = i \hat{P}_{01}^{(\mu)} - i \hat{P}_{10}^{(\mu)}
\nonumber\\
\hat{\lambda}_{3}^{(\mu)} &=& \hat{P}_{11}^{(\mu)} - \hat{P}_{00}^{(\mu)}
\,,\;\;\;\;
\hat{\lambda}_{0}^{(\mu)} = \hat{P}_{11}^{(\mu)} + \hat{P}_{00}^{(\mu)} = 
\hat{1}^{(\mu)}\,,
\end{eqnarray}
where $\hat{P}_{pq}^{(\mu)} = |p\rangle\!^{(\mu)} {}^{(\mu)}\hspace*{-0.8mm}
\langle q|$ is a (local) 
transition operator. For simplicity we restrict ourselves here to $M=1$. 

The initial state $|\psi_{0}\rangle$ will be taken to be 
a product of the Turing-head and tape wave-functions. 
For the discretized dynamical description of the QTM 
we identify the unitary operators $\hat{U}_{n},\,n=1,2,3,\cdots$ 
(step number) with the local unitary transformation on the Turing-head $S$, 
$\hat{U}_{\alpha_m}^{(S)}$, 
and the quantum-controlled-NOT (QCNOT) on ($S,1$), $\hat{U}^{(S,1)}$, 
respectively, as follows:
\begin{eqnarray}
&\hat{U}_{2m-1} = \hat{U}_{\alpha_{m}}^{(S)} = \hat{1}^{(S)}\;
\cos{(\alpha_{m}/2)} 
- \hat{\lambda}_{1}^{(S)}\;i\sin{(\alpha_{m}/2)}&\label{us}\\
&\hat{U}_{2m} = \hat{U}^{(S,1)} = 
\hat P_{00}^{(S)}\, \hat \lambda_1^{(1)} + \hat P_{11}^{(S)}\, \hat{1}^{(1)} = 
\left(\hat{U}^{(S,1)}\right)^{\dagger}&\,,\label{ub}
\end{eqnarray}
where the Turing head is externally driven by the Fibonacci-like sequence 
$\alpha_{m+1} = \alpha_{m} + \alpha_{m-1}$ (mod $2\pi$), 
$\alpha_0 = 0$. The $m$th Fibonacci number $\alpha_m$ is then controlled by 
$\alpha_1$ via
\begin{equation}
\label{fibonacci}
\alpha_m\, =\, \frac{\alpha_1}{\sqrt{5}} \left(\beta^{m}-\gamma^{m}\right)\,,
\end{equation}
where $\beta := \frac{1+\sqrt{5}}{2},\,\gamma := \frac{1-\sqrt{5}}{2}$. 
The sequence $\{\alpha_m\}$ (mod $2 \pi$) acts as a chaotic input 
(Lyapunov exponent: $\ln \beta > 0$). It is 
useful for later calculations to note that
\begin{eqnarray}
&&\beta^{m+1} = \beta^{m} + \beta^{m-1}\,,\;\;\;\;
\gamma^{m+1} = \gamma^{m} + \gamma^{m-1}\nonumber\\
&&\beta^{m} = F(m)\cdot \beta + F(m-1)\,,\;\;\;\;
\gamma^{m} = F(m)\cdot  \gamma + F(m-1)\,,
\label{rel_beta}
\end{eqnarray}
where $F(m) := \left(\beta^m - \gamma^m\right)/\sqrt{5}$, the $m$th Fibonacci 
number with $F(1) = 1$. The chaotic sequence of Fibonacci-type can be 
interpreted as a temporal random (chaotic) analogue to 
$1$-dimensional chaotic potentials in real space \cite{KOH83,PIE95}.

First, we restrict ourselves to the reduced state-space dynamics of 
the head~$S$ and tape-spin~$1$, respectively, 
\begin{equation}
\label{bloch}
\lambda_{i}^{(S)}(n) = \langle\psi_{n}|\hat{\lambda}_{i}^{(S)} \otimes 
\hat{1}^{(1)}
|\psi_{n}\rangle\,,\;\;\lambda_{k}^{(1)}(n) = 
\langle\psi_{n}|\hat{1}^{(S)} \otimes \hat{\lambda}_{k}^{(1)}|\psi_{n}
\rangle\,.
\end{equation}
$|\psi_n\rangle$ is the total network state at step $n$, 
$\lambda_{i}^{(\mu)}(n)$ are the respective Bloch-vectors. 
We intend to show that these local propagations are chaotic, too. 
Due to the entanglement between the head and tape, both will, in 
general, appear to be in a `mixed-state', which means that the length of the 
Bloch-vectors in (\ref{bloch}) is less than $1$. 
However, for specific initial states 
$|\psi_0\rangle$ the state of head and tape will remain pure: 
As $|\pm\rangle\!^{(1)} := \frac{1}{\sqrt{2}}\left(|0\rangle\!^{(1)} \pm 
|1\rangle\!^{(1)}\right)$ are the eigenstates of 
$\hat{\lambda}_{1}^{(1)}$ with $\hat{\lambda}_{1}^{(1)} 
|\pm\rangle\!^{(1)} = \pm |\pm\rangle\!^{(1)}$, 
the QCNOT-operation $\hat{U}^{(S,1)}$ of equation~(\ref{ub}) cannot create 
any entanglement, irrespective of the head state 
$|\varphi\rangle\!^{(S)}$, i.e.
\begin{eqnarray}
\label{entangle}
\hat{U}^{(S,1)}\,|\varphi\rangle\!^{(S)} \otimes\,|+\rangle\!^{(1)}\,&=&\,
|\varphi\rangle\!^{(S)} \otimes\,|+\rangle\!^{(1)}\nonumber\\
\hat{U}^{(S,1)}\,|\varphi\rangle\!^{(S)} \otimes\,|-\rangle\!^{(1)}\,&=&\,
\hat{\lambda}_{3}^{(S)} 
|\varphi\rangle\!^{(S)} \otimes\,|-\rangle\!^{(1)}\,.
\end{eqnarray}
As a consequence, the state~$|\psi_n\rangle$ remains a product state at any 
step~$n$ for the initial product states 
$|\psi_0^{\pm}\rangle = |\varphi_0\rangle\!^{(S)} \otimes 
|\pm\rangle\!^{(1)}$\, with 
$|\varphi_0\rangle\!^{(S)} = \cos{(\varphi_{0}/2)} |0\rangle\!^{(S)} - 
i \sin{(\varphi_{0}/2)} |1\rangle\!^{(S)}$ 
and the Turing head then performs a pure-state trajectory (`primitive') on 
the Bloch-circle $\left(\lambda_{1}^{(S)}(n)=0\right)$
\begin{equation}
\label{product}
|\psi_{n}^{\pm}\rangle = |\varphi_{n}^{\pm}\rangle\!^{(S)} \otimes 
|\pm\rangle\!^{(1)}\,,
\;\;\;\;\left(\lambda_{2}^{(S)}(n|\pm)\right)^{2} + 
\left(\lambda_{3}^{(S)}(n|\pm)\right)^{2} = 1\,.
\end{equation}

Here $\lambda_{j}^{(S)}(n|\pm)$ denotes the Bloch-vector of the 
Turing head $S$ conditioned by the initial state $|\psi_0^{\pm}\rangle$. 
From the Fibonacci relation and the property (\ref{entangle}) 
it is found for $|\varphi_{n}^{+}\rangle\!^{(S)} \otimes\,|+\rangle\!^{(1)}, 
n = 2m$, and $\varphi_0^{\pm} = \alpha_0 = 0$ that
\begin{equation}
\label{plus_fibo}
\lambda_{2}^{(S)}(2m|+) = \sin {\mathcal{C}}_{2m}(+)\,,\;\;\;\;
\lambda_{3}^{(S)}(2m|+) = -\cos {\mathcal{C}}_{2m}(+)\,,
\end{equation}
where\, ${\mathcal{C}}_{2m}(+) := 
{\displaystyle \sum_{j=1}^{m} \alpha_{j}}$\,, and 
$\lambda_{k}^{(S)}(2m-1|+) = \lambda_{k}^{(S)}(2m|+)$. In order to 
derive the corresponding expression of $\lambda_{k}^{(S)}(n|-)$ for 
$|\varphi_{n}^{-}\rangle\!^{(S)} \otimes\,|-\rangle\!^{(1)}$, 
we utilize the 
following recursion relations for the cumulative rotation angle 
${\mathcal{C}}_{n}(-)$ up to step $n$
\begin{equation}
\label{recursion_rel}
{\mathcal{C}}_{2m}(-) = -{\mathcal{C}}_{2m-1}(-)\,,\;\;\;\;
{\mathcal{C}}_{2m-1}(-) = \alpha_{m} + {\mathcal{C}}_{2m-2}(-)\,.
\end{equation}
Then ${\mathcal{C}}_{2m}(-), {\mathcal{C}}_{2m-1}(-)$, respectively, 
satisfy the following expressions:
\begin{eqnarray}
\label{cumulative}
{\mathcal{C}}_{2m}(-) &=& -{\mathcal{C}}_{2m-2}(-) - \alpha_{m}\, =\, 
(-1)^{m-1} \sum_{j=1}^{m}\,(-1)^{j} \alpha_{j}\nonumber\\
{\mathcal{C}}_{2m-1}(-) &=& -{\mathcal{C}}_{2m-3}(-) + \alpha_{m}\, =\, 
(-1)^{m} \sum_{j=1}^{m}\,(-1)^{j} \alpha_{j}\,,
\end{eqnarray}
yielding $\lambda_{2}^{(S)}(n|-) = \sin {\mathcal{C}}_{n}(-),\,
\lambda_{3}^{(S)}(n|-) = -\cos {\mathcal{C}}_{n}(-)$ 
$\left[\mbox{{\em cf.} (\ref{plus_fibo})}\right]$. 
The Fibonacci property implies that both primitives, 
$|\varphi_{n}^{+}\rangle\!^{(S)} \otimes\,|+\rangle\!^{(1)}$ and 
$|\varphi_{n}^{-}\rangle\!^{(S)} \otimes\,|-\rangle\!^{(1)}$, are 
chaotically driven. 

From any initial state, 
$|\psi_{0}\rangle = a^{(+)}|\varphi_{0}^{+}\rangle\!^{(S)} \otimes 
|+\rangle\!^{(1)} + a^{(-)}|\varphi_{0}^{-}\rangle\!^{(S)} \otimes 
|-\rangle\!^{(1)}$, we then obtain at step $n$
\begin{equation}
|\psi_n\rangle = a^{(+)}|\varphi_{n}^{+}\rangle\!^{(S)} \otimes\,
|+\rangle\!^{(1)}\,+\,a^{(-)}|\varphi_{n}^{-}\rangle\!^{(S)} \otimes\,
|-\rangle\!^{(1)}
\end{equation}
and, observing the orthogonality of the $|\pm\rangle\!^{(1)}$, 
\begin{equation}
\label{super}
\lambda_{k}^{(S)}(n) = |a^{(+)}|^{2}\,\lambda_{k}^{(S)} (n|+)\,+\,
|a^{(-)}|^{2}\,\lambda_{k}^{(S)} (n|-)\,.
\label{lambda_S}
\end{equation}
This trajectory of the Turing-head~$S$ represents a non-orthogonal pure-state 
decomposition. By using (\ref{plus_fibo}), (\ref{cumulative}), (\ref{super}) 
$\left( \mbox{with}\; a^{(+)} = a^{(-)} = 1/\sqrt{2} \right)$ 
we finally have for 
$|\psi_0\rangle = |0\rangle\!^{(S)} \otimes\,|0\rangle\!^{(1)}$
\begin{eqnarray}
\label{chaotic_driving}
\left(\lambda_{2}^{(S)}(2m),\,\lambda_{3}^{(S)}(2m)\right)\;&=&\;
\cos {\mathcal{A}}_m \cdot (\sin {\mathcal{B}}_m ,\,-\cos {\mathcal{B}}_m)
\nonumber\\
\left(\lambda_{2}^{(S)}(2m-1),\,\lambda_{3}^{(S)}(2m-1)\right)\;&=&\;
\cos {\mathcal{B}}_m \cdot (\sin {\mathcal{A}}_m ,\,-\cos {\mathcal{A}}_m)\,,
\label{cumul}
\end{eqnarray}
where ${\mathcal{A}}_m := \alpha_{m}+\alpha_{m-2}+\cdots\,$, 
${\mathcal{B}}_m := 
\alpha_{m-1}+\alpha_{m-3}+\cdots\,$. 
The expression~(\ref{chaotic_driving}) indicates that also in the 
`non-classical' regime the local dynamics of the Turing head is controlled 
by a `chaotic' driving force, since the 
sequences ${\mathcal{A}}_m$ and ${\mathcal{B}}_m$, namely $\{\alpha_{2m}\}$ 
(mod $2 \pi$) or $\{\alpha_{2m-1}\}$ (mod $2 \pi$), respectively, are in fact 
both chaotic, as $\{\alpha_{m}\}$ (mod $2 \pi$) is. 

The Bloch-vector 
$\vec{\lambda}^{(S)}(n)$ can alternatively be calculated 
directly from the initial state $\left( \mbox{here:}\, 
|\psi_0\rangle = |0\rangle\!^{(S)} \otimes\,|0\rangle\!^{(1)}\right)$ and 
for any control angle $\alpha_1$ by using equation~(\ref{cumul}) and 
the relations
\begin{eqnarray}
\label{AB}
{\mathcal{A}}_m &=& \left\{
\begin{array}{ll}
\frac{\alpha_1}{\sqrt{5}}\,\left(\beta^{m+1} - \gamma^{m+1}\right)&
\mbox{\hspace{2.8ex}}m = \mbox{odd}\\
\frac{\alpha_1}{\sqrt{5}}\,\left(\beta^{m+1} - \gamma^{m+1} - \sqrt{5}\right)&
\mbox{\hspace{2.8ex}}m = \mbox{even}
\end{array}
\right.\nonumber\\
{\mathcal{B}}_m &=& \left\{
\begin{array}{ll}
\frac{\alpha_1}{\sqrt{5}}\,\left(\beta^{m} - \gamma^{m} - \sqrt{5}\right)&
\mbox{\hspace{7ex}}m = \mbox{odd}\\
\frac{\alpha_1}{\sqrt{5}}\,\left(\beta^{m} - \gamma^{m}\right)&
\mbox{\hspace{7ex}}
m = \mbox{even}\,,
\end{array}\right.
\end{eqnarray}
demonstrating a striking computational reducibility in that one needs 
for calculating $\vec{\lambda}^{(S)}(n)$ neither the total network 
state $|\psi_n\rangle$ nor to follow up each individual step $n$.

\section{Instability of periodic orbits}

Now we verify that the periodic orbits on the plane 
$\left\{0, \lambda_{2}^{(S)}, 
\lambda_{3}^{(S)}\right\}$ are unstable, which proves 
that the dynamics of the Turing head (`output') is indeed chaotic 
(figure~\ref{QTM_chaos}). 
Because of the alternating character of the dynamics, 
equations~(\ref{us}), (\ref{ub}), 
it suffices to check the periodicity only for 
step $n=2m$: The periodic orbits for 
$\lambda_{2}^{(S)}(0) = 0, \lambda_{3}^{(S)}(0) = -1$ 
must obey two constraints, 
${\mathcal{C}}_{2m}(+) = {\mathcal{C}}_{2m}(-) \stackrel{!}{=} 
2 {\pi} p,\,p \in \mathbf{Z}$ and $\alpha_{m+1} = \alpha_{1}$ (mod $2 \pi$) 
for $n = 2m+1$ 
(one concludes that $\alpha_1$ must be a rational multiple of $\pi$). 
By using the Fibonacci numbers (\ref{fibonacci}), one finds 
${\mathcal{C}}_{2m}^{\mbox{per}}(+)$ in (\ref{plus_fibo}) and 
${\mathcal{C}}_{2m}^{\mbox{per}}(-)$ in (\ref{cumulative}), respectively, 
for period $= 2m$ as
\begin{eqnarray}
\label{plus_po}
{\mathcal{C}}_{2m}^{\mbox{per}}(+)\, &=&\, \frac{\alpha_1}{\sqrt{5}}\, 
\left(\beta^{m+2} - \gamma^{m+2} - \sqrt{5} \right)\nonumber\\
{\mathcal{C}}_{2m}^{\mbox{per}}(-)\, &=&\, \frac{\alpha_1}{\sqrt{5}}\, 
\left(-\beta^{m-1} + \gamma^{m-1} + (-1)^{m} \sqrt{5} \right)\,.
\end{eqnarray}
Then let us consider a small perturbation $\delta$ of the initial 
phase angle $\alpha_0 = 0$, implying 
$|\varphi_0\rangle\!^{(S)} = \cos(\delta/2)|0\rangle - 
i \sin(\delta/2)|1\rangle$ 
and a perturbed Fibonacci-like sequence $\{\alpha_m'\}$ 
(mod $2 \pi$):
\begin{equation}
\alpha_{0}' = \delta,\, \alpha_{1}' = \alpha_1,\, \alpha_{2}' = \alpha_1 + 
\delta,\, \cdots\,.
\label{pertur}
\end{equation}
Similarly to (\ref{plus_po}), we obtain for this case 
${\mathcal{C}}_{2m}'(\pm) = {\mathcal{C}}_{2m}^{\mbox{per}}(\pm) + 
\Delta {\mathcal{C}}_{2m}(\pm)$, where the deviation terms from the periodic 
orbits read, respectively,
\begin{eqnarray}
\label{pm_per}
\Delta {\mathcal{C}}_{2m}(+) &=& \frac{\delta}{\sqrt{5}} 
\left(\beta^{m+1} - \gamma^{m+1}\right)\nonumber\\
\Delta {\mathcal{C}}_{2m}(-) &=& -\frac{\delta}{\sqrt{5}} 
\left(\beta^{m-2} - \gamma^{m-2}\right)\,.
\end{eqnarray}
By using (\ref{pm_per}), we are able to represent the evolution of the 
perturbation, $\Delta \lambda_{j}^{(S)}(n)$, 
at the $n=2m$-th step for the Turing-head dynamics as
\begin{equation}
\left( \begin{array}{c}
        \Delta \lambda_{2}^{(S)}(2m)\\
        \Delta \lambda_{3}^{(S)}(2m)
        \end{array} \right) = \left( \begin{array}{cc}
                                     M_{11} & 0\\
                                     0 & M_{22}
                                     \end{array} \right) 
                              \left( \begin{array}{c} 
                                    \Delta \lambda_{2}^{(S)}(0)\\
                                    \Delta \lambda_{3}^{(S)}(0)
                                    \end{array} \right)\,,
\end{equation}
where $\Delta \lambda_{2}^{(S)}(0)=\sin \delta,\,\Delta 
\lambda_{3}^{(S)}(0)=-\cos \delta;\,\Delta \lambda_{2}^{(S)}(2m)=
\cos(\delta \alpha_m) \sin$ 
$(\delta \alpha_{m-1}),\,\Delta \lambda_{3}^{(S)}(2m)$\linebreak
$=-\cos(\delta \alpha_m) \cos(\delta \alpha_{m-1});\,M_{11}=
\cos(\delta \alpha_{m}) \sin(\delta \alpha_{m-1})$ 
$/\sin \delta,\,M_{22}=\cos(\delta \alpha_{m}) 
\cos(\delta \alpha_{m-1})/$\linebreak
$\cos \delta$, respectively. 
One easily shows $\left[\mbox{{\em cf.} equation~(\ref{rel_beta})}\right]$ that
\begin{equation}
\lim_{\delta \to 0} M_{11} = F(m-1) = \frac{1}{\sqrt{5}} 
\left(\beta^{m-1} - \gamma^{m-1}\right)
\,\;\;\;\;\lim_{\delta \to 0} M_{22} = 1\,,
\end{equation}
which means that $M_{11}$ grows exponentially (note that 
$\beta > 1, |\gamma| < 1$), 
and the periodic orbit on the Turing head 
is thus unstable to any small perturbation $\delta$ in the external control. 

As an explicit example consider 
the case of $\alpha_1 = \frac{2 \pi}{5}$ with 
$|\psi_0\rangle = |0\rangle\!^{(S)} \otimes |0\rangle\!^{(1)}$: 
By using equation~(\ref{rel_beta}) in equation~(\ref{plus_po}) for 
${\mathcal{C}}_{2m}^{\mbox{per}}(+) = 2 {\pi} p$ the periodic orbits 
with period $=2m$ have to obey $F(m+2) = 1$ (mod 5). 
Together with the second condition, $F(m+1) = 1$ (mod 5) for 
$n = 2m+1$, we obtain $F(m-1) = 1$, $F(m) = 0$ (mod 5). For 
${\mathcal{C}}_{2m}^{\mbox{per}}(-) = 2 {\pi} p$ it follows likewise that 
$F(m-1) = 1$ (mod 5) for $m =$ even, 
$F(m-1) = 4$ (mod 5) for $m =$ odd. Considering the common condition for 
both cases, $F(m-1) = 1$ and $F(m) = 0$ (mod 5), we find the smallest $m$ 
satisfying both conditions, namely $m = 20$, i.e. $n = 2 m = 40$. 
$\left[\displaystyle{\lim_{\delta \to 0} M_{11} = F(19) = 4181 \gg 1}, 
\mbox{see figure~\ref{stability}}\right]$.

Remarkably enough, the local dynamics of the Turing tape also contains 
some exponential sensitivity to initial conditions 
$\left[ \mbox{here:}\, |\psi_0\rangle = |0\rangle\!^{(S)} \otimes 
|0\rangle\!^{(1)}, \lambda_1^{(1)}(n) = \lambda_2^{(1)}(n) = 0\right]$:
\begin{equation}
\lambda_{3}^{(1)}(n) = \left\{
\begin{array}{cl}
-\cos\left(\alpha_{\left[\frac{n}{2}\right]+1}\, -\, \alpha_1\, +\, 
\delta_{\left[\frac{n}{2}\right]}^{\mbox{Fib}}\right)&\;\;\;\;n=0,1\;
(\mbox{mod}\; 4)\\
&\\
\cos\left(\alpha_{\left[\frac{n}{2}\right]+1}\, +\, 
\delta_{\left[\frac{n}{2}\right]}^{\mbox{Fib}}\right)&\;\;\;\;n=2,3\;
(\mbox{mod}\; 4)\,,
\end{array}
\right.
\end{equation}
where $\delta_{m}^{\mbox{Fib}} := \frac{\delta}{\sqrt{5}} 
(\beta^m - \gamma^m);\;[a] := n,\,a = n + r,\,n \in {\mathbf{Z}},\,
0 \leq r < 1$. One easily confirms that there is no periodic orbit with 
period $2m = 2$ (mod\,$4$). The following expression for the deviation from 
the periodic orbits with period $2m = 0$ 
(mod $4$) at step $n = 2m+2$ holds:
\begin{equation}
\Delta \lambda_{3}^{(1)}(2m+2) = M \cdot \Delta \lambda_{3}^{(1)}(2)
\,,\;\;\;\;
\lim_{\delta \to 0} M = \frac{1}{\sqrt{5}} \left(\beta^{m+1} - 
\gamma^{m+1}\right)\,
\frac{\sin(\alpha_{m+2})}{\sin(\alpha_{1})}\,,
\label{chaos_swapping}
\end{equation}
with $\Delta \lambda_{3}^{(1)}(2) = \cos(\alpha_1 + \delta) - 
\cos(\alpha_1),\,
\Delta \lambda_{3}^{(1)}(2m+2) = \cos(\alpha_{m+2} + \delta_{m+1}) - 
\cos(\alpha_{m+2})$; the perturbation $\delta$ in (\ref{pertur}) 
does not enter into the Turing-tape dynamics until $n = 2$\hspace{0.2ex}! 
Equation~(\ref{chaos_swapping}) 
shows the exponential instability of the periodic 
orbit. Indeed, the Turing tape exhibits chaos only by means of the 
entanglement with the head (`chaos swapping'), not as a result of a direct 
chaotic driving force. 

\section{Exponential parameter sensitivity}

The distance between density operators, $\hat{\rho}$ and $\hat{\rho}'$, 
defined by the so-called Bures metric \cite{HUE92}
\begin{equation}
D_{\rho \rho'}^{2} := 
\mbox{Tr} \left\{(\hat{\rho} - \hat{\rho}')^2\right\}\,.
\end{equation}
lies, independent of the dimension of the Liouville space, between 
0 and 2 [\,the maximum (squared) distance of 2 applies to pure 
orthogonal states, $D^2 = 2\,(1 - |\langle\psi|\psi'\rangle|^2)$\,]. 
This metric can be applied to the total-network-state space 
or any subspace. In any case it is a convenient additional means to 
characterize various QTMs: For $\alpha_m = \alpha_1$ (Lyapunov exponent $= 0$) 
\cite{KIM99} and any $\delta$ 
the distance remains almost constant (see figure~\ref{2figs}$d$); 
for the Fibonacci-like driving force, on the other hand, $\left( \alpha_m 
(\hat{\rho}) = \alpha_m, \alpha_m (\hat{\rho}') = \alpha_m (\hat{\rho}) + 
\delta_{m-1}^{\mbox{Fib}} \right)$ 
we obtain an initial exponential sensitivity, 
which is eventually constrained, though, by $D^2 \leq 2$ 
(see figure~\ref{2figs}$a$ - $c$). 
It is instructive to consider another regular QTM controlled by the rule 
$\alpha_{m+1} = 2 \alpha_{m} - \alpha_{m-1}$ (Lyapunov exponent $= 0$): 
$\alpha_m (\hat{\rho}) = m \alpha_1, \alpha_m (\hat{\rho}') = 
\alpha_m (\hat{\rho}) - (m-1)\,\delta$. 
Here we observe a revival in the evolution of $D^2$, which confirms that 
periodic orbits are stable (figure \ref{2figs}$d$), 
whereas there is no revival for the above chaotic system. Finally we display 
the evolution of $D^2$ for the total network state $|\psi_n\rangle$, which 
also shows exponential sensitivity. The respective distances 
for tape-spin $1$ are similar to those shown.

The ultimate source of the present chaotic behaviour is that any small 
perturbation $\delta$ to the initial state, $|\psi_0(\delta)\rangle$, 
is connected with a perturbed 
unitary evolution, $\hat{U}(\delta)$, which implies that the scalar product 
between different initial states 
(as a measure of distance) is no longer conserved under these evolutions:
\begin{equation}
O' := |\langle \psi_0(\delta)| \hat{U}^{\dagger}(\delta)\, 
\hat{U}(0) |\psi_0(0) \rangle|^2\,,\;\;\;\;D^2 = 2 ( 1 - O')\,.
\end{equation}
Thus the initial state is directly correlated to its unitary evolution, which 
can lead to the exponential sensitivity to initial condition, 
whereas there is no chaos in a generic quantum system evolving by a fixed 
$\hat{U}$ even if characterized by chaotic input parameters. 
This $O'$ reminds us immediately of the test function 
$O = |\langle \psi| \hat{V}^{\dagger}(t)\, \hat{U}(t) 
|\psi \rangle|^2$ \cite{PER91}, where $\hat{U}, \hat{V}$ are specified by 
slightly different external parameters: The corresponding 
parameter-sensitivity has been proposed as a measure to 
distinguish quantum chaos from regular quantum dynamics. 
The origin of chaos in our QTM may thus be alternatively ascribed to a 
perturbed $\hat{V} = \hat{U}(\delta)$ in the control ({\em cf.} the comment 
by R.~Schack \cite{SCH95}).

\section{Summary}

In conclusion, we have studied the quantum dynamics of a small chaotically 
driven QTM based on a decoherence-free Hamiltonian. 
Quantum chaos has been shown to occur as an exponential parameter-sensitivity 
and a cumulative loss of control in a pure quantum regime. 
This might be contrasted with the usual 
quantum chaology, which is concerned essentially with 
semiclassical spectrum analysis of classically chaotic systems 
(e.g. level spacing, spectral rigidity). As quantum features 
we utilized the superposition principle and the physics of entanglement. 
Our dynamical chaos manifests itself in the superposition and entanglement of 
a pair of `classical' (i.e. unentangled) chaotic state-sequences. 
Due to the entanglement, we can see the 
chaos in any local Bloch-plane. This indicates that patterns in reduced 
Bloch-spheres (a quantum version of a 
Poincar\'{e}-cut, figure~\ref{stability}) should be useful to 
characterize quantum chaos in a broad class of quantum networks: Here, 
a periodic orbit would be represented by finite set of fixed points on the 
plane $\left\{0, \lambda_{2}^{(S)}, \lambda_{3}^{(S)}\right\}$. 
It is noteworthy that this kind of control loss is completely different from 
the typical control limit of a quantum network resulting from the 
exponential blow-up of Hilbert-space dimension in which the state evolves 
\cite{FEY82}. It is expected that a QTM architecture with a larger number of 
pseudo-spins on the Turing tape would also exhibit chaos 
under the same type of driving. However, it is just the chaos in small 
networks which might be interesting for experimental studies, especially in 
the form of ensembles thereof. The Fibonacci-like sequence should, however, be 
considered but a special example for chaotic input. Such inputs 
would, of course, have to be avoided in quantum computation; otherwise the 
resulting quantum dynamics would easily become chaotic in its entirety!

\section{Acknowledgements}

We thank C.~Granzow, M.~Karremann, A.~Otte and, P.~Pangritz 
for fruitful discussions, and especially P.~Pangritz for numerical results. 
One of us (I.~K.) acknowledges D.~Braun and S.~Gnutzmann for useful references.

\begin{figure}
\refstepcounter{figure}\label{QTM_chaos}
\refstepcounter{figure}\label{stability}
\refstepcounter{figure}\label{2figs}
\vspace*{25.5cm}
\hspace*{-1.55cm}
\includegraphics{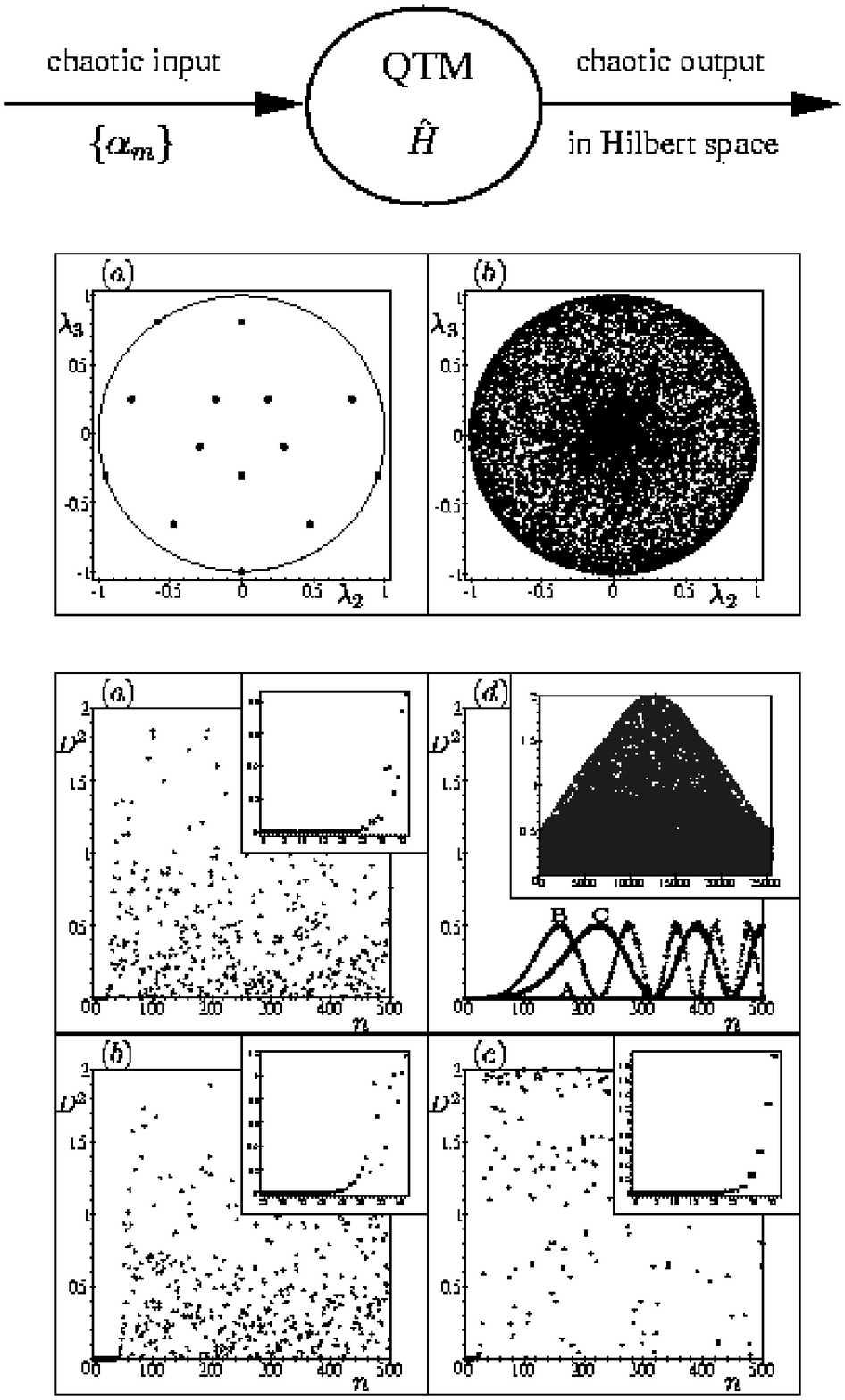}
\end{figure}

\newpage
Figure~\ref{QTM_chaos}: 
A input-output scheme of our quantum Turing machine~(QTM).
\vspace*{0.5cm}

Figure~\ref{stability}: Turing-head patterns 
$\left\{0, \lambda_{2}(n), \lambda_{3}(n)\right\}$ 
under Fibonacci control for initial state 
$|\psi_{0}\rangle = 
|0\rangle\!^{(S)} \otimes\,|0\rangle\!^{(1)}$. ($a$): $\alpha_1 = 
\frac{2}{5} \pi$ (periodic);\, ($b$): $\alpha_1 = 
\frac{2}{5} \times 3.141592654$ (aperiodic) and total step number $n=10000$.
\vspace*{0.5cm}

Figure~\ref{2figs}: Evolution of the (squared) distance $D_{\rho \rho'}^2$ 
between QTM state with $\left(\hat{\rho}'\right)$ and 
without $\left(\hat{\rho}\right)$ perturbation $\delta$. 
$\alpha_1 = \frac{2}{5} \pi$, 
$|\psi_{0}\rangle = |0\rangle\!^{(S)} \otimes\,|0\rangle\!^{(1)}$ for 
$\hat{\rho}$, and $\left(\cos(\delta/2)|0\rangle\!^{(S)}\, -\right.$\linebreak
$\left.i \sin(\delta/2) |1\rangle\!^{(S)}\right) 
\otimes\,|0\rangle\!^{(1)}$ for $\hat{\rho}'$. ($a$): 
chaotic input according to 
equation~(\ref{fibonacci}) (inset shows initial behaviour in more detail; 
{\em cf.} Table~\ref{chaos_table}) for the 
Turing-head state, $\delta = 0.001$;\, ($b$): the same as ($a$) but for 
$\delta = 0.00001$;\, 
($c$): the same as ($a$) for total network state $|\psi_n\rangle$;\, 
($d$): $D_{\rho \rho'}^2$ within the Turing head subspace for 
$\alpha_m = \alpha_1$ ($D^2 \approx 0$, solid line A, $\delta = 
0.001, 0.0005$) and $\alpha_{m+1} = 2 \alpha_{m} - \alpha_{m-1}$ 
(dotted line B, $\delta = 0.001$; boxed line C, $\delta = 0.0005$), inset 
shows line B on larger scale.
\begin{table}
\caption{Analytical evolution of (squared) distance $D_{\rho \rho'}^2$ 
between the Turing-head states as of Fig~\ref{2figs}$a$: 
With increasing time step $n$ the originally small phase angle 
grows at a Fibonacci-like rate (separately for $n = $ even and $n = $ odd), 
such that $D^2$ shows exponential growth initially and then wild oscillations 
between 0 and 2. \label{chaos_table}}
\vspace*{1cm}
\begin{tabular}{c|l}
\hline
\hline
$n$& $D_{\rho \rho'}^2(n)$\\
\hline
\hline
$0$& $1 - {\rm cos}({\mathbf{1\delta}})$\\
\hline
\hline
$1$& $1 - {\rm cos}({\mathbf{1\delta}})$\\
\hline
\hline
$2$& $\frac{1}{2} + \frac{1}{4} {\rm cos}(2\alpha_1 + {\mathbf{2\delta}}) - 
\frac{1}{2} {\rm cos}(2\alpha_1 + {\mathbf{1\delta}}) - \frac{1}{2} 
{\rm cos}({\mathbf{1\delta}}) + \frac{1}{4} {\rm cos}(2\alpha_1)$\\
\hline
\hline
$3$& $\frac{1}{4} - \frac{1}{4} {\rm cos}({\mathbf{2\delta}})$\\
\hline
\hline
$4$& $\frac{1}{4} - \frac{1}{4} {\rm cos}({\mathbf{2\delta}})$\\
\hline
\hline
$5$& $\frac{1}{2} - \frac{1}{4} {\rm cos}(2\alpha_1 + {\mathbf{3\delta}}) 
- \frac{1}{4} {\rm cos}({\mathbf{3\delta}}) + \frac{1}{4} 
{\rm cos}(2\alpha_1 + {\mathbf{2\delta}})$\\
& $- \frac{1}{4} {\rm cos}(2\alpha_1 - {\mathbf{1\delta}}) - \frac{1}{4} 
{\rm cos}({\mathbf{1\delta}}) + \frac{1}{4} {\rm cos}(2\alpha_1)$\\
\hline
\hline
$6$& $\frac{1}{2} + \frac{1}{4} {\rm cos}(6\alpha_1 + {\mathbf{4\delta}}) - 
\frac{1}{4} {\rm cos}(6\alpha_1 + {\mathbf{3\delta}}) - \frac{1}{4} 
{\rm cos}({\mathbf{3\delta}})$\\
& $- \frac{1}{4} {\rm cos}(6\alpha_1 + {\mathbf{1\delta}}) - 
\frac{1}{4} {\rm cos}({\mathbf{1\delta}}) + \frac{1}{4} {\rm cos}(6\alpha_1)$\\
\hline
\hline
$7$& $\frac{1}{2} - \frac{1}{4} {\rm cos}(6\alpha_1 + {\mathbf{5\delta}}) 
- \frac{1}{4} {\rm cos}({\mathbf{5\delta}}) + \frac{1}{4} {\rm cos} 
(6\alpha_1 + {\mathbf{4\delta}})$\\
& $- \frac{1}{4} {\rm cos}(6\alpha_1 - {\mathbf{1\delta}}) - \frac{1}{4} 
{\rm cos}({\mathbf{1\delta}}) + \frac{1}{4} {\rm cos}(6\alpha_1)$\\
\hline
\hline
$8$& $\frac{1}{2} + \frac{1}{4} {\rm cos}(8\alpha_1 + {\mathbf{6\delta}}) - 
\frac{1}{4} {\rm cos}(8\alpha_1 + {\mathbf{5\delta}})  - 
\frac{1}{4} {\rm cos}({\mathbf{5\delta}})$\\
& $- \frac{1}{4} {\rm cos}(8\alpha_1 + {\mathbf{1\delta}}) - 
\frac{1}{4}{\rm cos}({\mathbf{1\delta}}) + \frac{1}{4} {\rm cos}(8\alpha_1)$\\
\hline
\hline
$9$& $\frac{1}{2}  - \frac{1}{4} {\rm cos}(8\alpha_1 + {\mathbf{8\delta}}) 
- \frac{1}{4} {\rm cos}({\mathbf{8\delta}}) + \frac{1}{4} {\rm cos}
(8\alpha_1 + {\mathbf{6\delta}})$\\
& $- \frac{1}{4} {\rm cos}(8\alpha_1 - {\mathbf{2\delta}}) - \frac{1}{4} 
{\rm cos}({\mathbf{2\delta}}) + \frac{1}{4} {\rm cos}(8\alpha_1)$\\
\hline
\hline
$10$& $\frac{1}{2} + \frac{1}{4} {\rm cos}(16\alpha_1 + {\mathbf{10\delta}}) 
- \frac{1}{4} {\rm cos}(16\alpha_1 + {\mathbf{8\delta}}) - \frac{1}{4} 
{\rm cos}({\mathbf{8\delta}})$\\
& $- \frac{1}{4} {\rm cos}(16\alpha_1 + {\mathbf{2\delta}}) - \frac{1}{4} 
{\rm cos}({\mathbf{2\delta}}) + \frac{1}{4} {\rm cos}(16\alpha_1)$\\
\hline
\hline
$11$& $\frac{1}{2} - \frac{1}{4} {\rm cos}(16\alpha_1 + {\mathbf{13\delta}}) 
- \frac{1}{4} {\rm cos}({\mathbf{13\delta}}) + 
\frac{1}{4} {\rm cos}(16\alpha_1 + {\mathbf{10\delta}})$\\
& $- \frac{1}{4} {\rm cos}(16\alpha_1 - {\mathbf{3\delta}}) - \frac{1}{4} 
{\rm cos}({\mathbf{3\delta}}) + \frac{1}{4} {\rm cos}(16\alpha_1)$\\
\hline
\hline
$12$& $\frac{1}{2} + \frac{1}{4} {\rm cos}(24\alpha_1 + {\mathbf{16\delta}}) 
- \frac{1}{4} {\rm cos}(24\alpha_1 + {\mathbf{13\delta}}) - \frac{1}{4} 
{\rm cos}({\mathbf{13\delta}})$\\
& $- \frac{1}{4} {\rm cos}(24\alpha_1 + {\mathbf{3\delta}}) - \frac{1}{4} 
{\rm cos}({\mathbf{3\delta}}) + \frac{1}{4} {\rm cos}(24\alpha_1)$\\
\hline
\hline
\end{tabular}
\end{table}

\begin{thebibliography}{99}
\bibitem{SCH35} E.~Schr\"{o}dinger, 1935, 
{\em Naturwissenschaften\/}, {\bf 23}, 807, 823 and 844, reprinted in 
English in: J.~A.~Wheeler and W.~H.~Zurek, 1983, {\em Quantum Theory of 
Measurement\/} (Princeton U.~P., Princeton, N.~J.).

\bibitem{DUE98} S.~D\"{u}rr, T.~Nonn, and G.~Rempe, 1998, 
{\em Nature\/}, {\bf 395}, 33.

\bibitem{BER85} M.~V.~Berry, 1985, 
{\em Proc.~R.~Soc.~London~A\/}, {\bf 400}, 229; 1987, 
{\em Proc.~R.~Soc.~London~A\/}, {\bf 413}, 
183; 1989, {\em Physica Scripta\/} {\bf 40}, 335.

\bibitem{FOR91} J.~Ford, G.~Mantica, and G.~H.~Ristow, 1991, 
{\em Physica~D\/}, {\bf 50}, 493.

\bibitem{HAA91} F.~Haake, 1991, 
{\em Quantum Signatures of Chaos\/} (Springer, New York).

\bibitem{KOU97} L.~P.~Kouwenhoven {\em et al.}, 1997, 
in {\em Mesoscopic Electron Transport\/}, edited by L.~L.~Sohn {\em et al.}, 
NATO ASI Series E345 (Kluwer, Dordrecht).

\bibitem{SHE94} D.~L.~Shepelyansky, 1994, 
{\em Phys.~Rev.~Lett.\/}, {\bf 73}, 2607.

\bibitem{WAI98} X.~Waintal and J-L.~Pichard, 1998, 
{\em Eur.~Phys.~J.~B\/}, {\bf 6}, 117; X.~Waintal, D.~Weinmann, and 
J-L.~Pichard, 1999, {\em Eur.~Phys.~J.~B\/}, {\bf 7}, 451.

\bibitem{BEN82} P.~Benioff, 1982, 
{\em Phys.~Rev.~Lett.\/}, {\bf 48}, 1581; 1996, {\em Phys.~Rev.~A\/}, 
{\bf 54}, 1106; 1998, {\em Fortschr.} {\em Physik\/}, {\bf 46}, 423.

\bibitem{DEU85} D.~Deutsch, 1985, 
{\em Proc.~R.~Soc.~London~A\/}, {\bf 400}, 97; 1989, 
{\em Proc.~R.~Soc.~London~A\/}, {\bf 425}, 73.

\bibitem{SCH98} R.~Schack, 1998, 
{\em Phys.~Rev.~A\/}, {\bf 57}, 1634; 
T.~Brun and R.~Schack, quant-ph/9807050.

\bibitem{GAR97} S.~A.~Gardiner, J.~I.~Cirac, and P.~Zoller, 1997, 
{\em Phys.~Rev.~Lett.\/}, {\bf 79}, 4790; 
Erratum, 1998, {\em Phys.~Rev.~Lett.\/}, {\bf 80}, 2968.

\bibitem{BER99} G.~P.~Bergman {\em et al.}, quant-ph/9903063.

\bibitem{GHA97} M.~E.~Ghafar {\em et al.}, 1997, {\em J.~Mod.~Optics\/}, 
{\bf 44}, 1985; 1997, {\em Phys.~Rev.~Lett.\/}, {\bf 78}, 4181.

\bibitem{RIE99} K.~Riedel {\em et al.}, 1999, {\em Phys.~Rev.~A\/}, {\bf 59}, 
797.

\bibitem{BLU94} R.~Bl\"{u}mel, 1994, {\em Phys.~Rev.~Lett.\/}, {\bf 73}, 428.

\bibitem{KIM99} I.~Kim and G.~Mahler, 1999, {\em Phys.~Rev.~A\/}, 
{\bf 60}, 692.

\bibitem{MAH98} G.~Mahler and V.~A.~Weberru{\ss}, 1998, 
{\em Quantum Networks: Dynamics of Open Nanostructures\/} 
(2nd ed. Springer, New York).

\bibitem{KOH83} M.~Kohmoto, L.~P.~Kadanoff, and C.~Tang, 1983, 
{\em Phys.~Rev.~Lett.\/}, {\bf 50}, 1870.

\bibitem{PIE95} F.~Pi\'{e}chon, M.~Benakli, and A.~Jagannathan, 1995, 
{\em Phys.~Rev.~Lett.\/}, {\bf 74}, 5248.

\bibitem{HUE92} M.~H\"{u}bner, 1992, 
{\em Phys.~Lett.~A\/}, {\bf 163}, 293; 
1993, {\em Phys.~Lett.~A\/}, {\bf 179}, 226.

\bibitem{PER91} A.~Peres, 1991, 
in {\em Quantum Chaos\/}, edited by H.~A.~Cerdeira {\em et al.} 
(World Scientific, Singapore); 1993, 
{\em Quantum Theory: Concepts and Methods\/} (Kluwer, Dordrecht).

\bibitem{SCH95} R.~Schack, 1995, 
{\em Phys.~Rev.~Lett.\/}, {\bf 75}, 581.

\bibitem{FEY82} R.~P.~Feynman, 1982, 
{\em Int. J. theor. Phys.\/}, {\bf 21}, 467.
\end{thebibliography}
\end{document}